\documentclass[onecolumn,notitlepage, prd,preprintnumbers,superscriptaddress,longbibliography,fleqn,nofootinbib]{revtex4-2}

\usepackage{amsmath,amssymb,graphicx,booktabs,bm,psfrag,color,slashed}
\usepackage[utf8]{inputenc}
\usepackage[colorlinks=true,linkcolor=blue,citecolor=blue]{hyperref}
\usepackage[compat=1.0.0]{tikz-feynman}
\usepackage{comment}

\newcommand{\nn}{\nonumber}

\newcommand{\be}{\begin{equation}}
\newcommand{\ee}{\end{equation}}
\newcommand{\bea}{\begin{eqnarray}}
\newcommand{\eea}{\end{eqnarray}}
\newcommand{\balign}{\begin{align}}
\newcommand{\ealign}{\end{align}}
\newcommand{\as}{\alpha_s}

\newcommand{\bg}{\begin{gather}}
\newcommand{\foma}{\end{gather}}

\newcommand{\noopsort}[1]{}

\def\<{\langle}
\def\>{\rangle}

\def\a{\alpha}

\def\g{\gamma}  
  \def\D{\Delta}

\def\({\left(}
\def\[{\left[}
\def\){\right)}
\def\]{\right]}

\def\ln{\hbox{ln}}

\def\le{\left }
\def\ri{\right}

\newcommand{\ben}{\begin{eqnarray}}
\newcommand{\een}{\end{eqnarray}}

\newcommand{\bef}{\begin{figure}[htb]\centering}
\newcommand{\eef}{\end{figure}}

\usepackage[normalem]{ulem} 
\renewcommand\sout{\bgroup \color[rgb]{1,0,0} \ULdepth=-.5ex \ULset}

\begin{document}

\title{Illuminating the nucleon spin}


\author{Miguel G. Echevarria}
\email{miguel.garciae@ehu.eus}
\affiliation{Department of Physics and EHU Quantum Center, University of the Basque Country UPV/EHU,\\ P.O. Box 644, 48080 Bilbao, Spain}

\begin{abstract}
In this short note the QED extension of the nucleon spin sum rule is considered.
To this end, the leading-order (LO) QED evolution kernels for the quark/gluon helicity and orbital-angular-momentum (OAM) distributions are calculated, as well as their lepton and photon analogue distributions introduced for the first time.
The LO evolution kernels of the latter are also calculated, both in QCD and QED.
Putting all together, the nucleon spin sum rule remains scale-invariant in QCD$\times$QED, as expected, which represents a check of the newly obtained results.
This theoretical development will allow in the future the quantification of the contributions of lepton and photon distributions to the nucleon spin, and a more precise control over the uncertainties.

\end{abstract}

\maketitle

\section{Introduction}

The Jaffe-Manohar \cite{Jaffe:1989jz} spin sum rule for the nucleon is 
\begin{equation}
\label{eq:jm}
\frac{1}{2}\Delta\Sigma(Q^2) 
+ \Delta G(Q^2) 
+ L_q(Q^2) 
+ L_g(Q^2) 
= \frac{1}{2}
\,,
\end{equation}
where the involved terms are given by the first moments of their corresponding partonic quark/gluon helicity and orbital angular momentum (OAM) distributions:
\begin{align}
\label{eq:definitionsQCD}
\Delta\Sigma(Q^2) &= 
\sum_{f}\int_0^1dx \ (\Delta q_f(x,Q^2)+\Delta \bar{q}_f(x,Q^2))
\,,\nn\\
\Delta G(Q^2)&=
\int_0^1dx \ \Delta G(x,Q^2)
\,,\nn\\
L_{q}(Q^2)&=
\sum_f \int_0^1dx \ (L_{f}(x,Q^2)+\bar{L}_f(x,Q^2))
\,,\nn\\
L_g(Q^2)&=
\int_0^1 dx \ L_g(x,Q^2)
\,.
\end{align}
As indicated in \eqref{eq:jm}, the four terms depend on the scale, although this dependence is obviously cancelled in the sum.
The QCD evolution kernels of the quark and gluon helicities were obtained at leading order (LO) in~\cite{Ahmed:1976ee,Altarelli:1977zs}, at next-to-leading order (NLO) 
in~\cite{Kodaira:1979pa,Mertig:1995ny,Vogelsang:1995vh,Vogelsang:1996im}, 
and more recently at next-to-next-to-leading order (NNLO) in~\cite{Vogt:2008yw,Moch:2014sna,Moch:2015usa}.
On the other hand, the QCD evolution kernels for the quark and gluon OAM distributions are known only at LO~\cite{Ji:1995cu,Hagler:1998kg,Harindranath:1998ve}.

The QCD evolution of the distributions is given by the coupled equations (see e.g. \cite{Hatta:2018itc} and references therein~\footnote{Notice that the splitting kernels are denoted differently as compared to \cite{Hatta:2018itc}.})
\begin{align}
\frac{d}{d\ln Q^2} \left(\begin{matrix} 
\Delta \Sigma (Q^2) \\ 
\Delta G (Q^2) \\
L_q(Q^2) \\
L_g(Q^2)
\end{matrix}\right) = 
\int^1_0 dx \int_x^1 \frac{dz}{z} 
\left(\begin{matrix} 
\Delta P(z,Q^2) &  \D\hat P(z,Q^2) \\  
\Omega \hat P(z,Q^2) & \Omega P(z,Q^2) 
\end{matrix}\right) 
\left(\begin{matrix}   
\Delta \Sigma (\tfrac{x}{z},Q^2) \\ 
\Delta G (\tfrac{x}{z},Q^2) \\
L_q(\tfrac{x}{z},Q^2) \\
L_g(\tfrac{x}{z},Q^2)
\end{matrix}\right)
\,,
\end{align}
with the elements of the evolution-kernel  matrix being each a $2\times2$ matrix like
\begin{align}
\Delta P(z,Q^2) = 
\left(\begin{matrix} 
\Delta P_{qq}(z,Q^2) &  \Delta P_{qg}(z,Q^2) \\  
\Delta P_{gq}(z,Q^2) & \Delta P_{gg}(z,Q^2) \end{matrix}\right)
\,,
\end{align}
and similarly for the rest.
The matrix $\Delta\hat P=0$ to all orders, since the evolution of the helicity distributions is governed solely by themselves with no mixing with the OAM operators.

Let us now briefly review the scale-invariance of the spin sum rule at LO in QCD.
For that we need the perturbative expansions of the distributions and the evolution kernels, which we denote as
\begin{align}
\label{eq:expansionsQCD}
& \Delta \Sigma (x,Q^2)=2n_f\delta(1-x) 
+ \sum_{n=1} \Big(\frac{\alpha_s(Q^2)}{2\pi}\Big)^n\Delta \Sigma^{(n)} (x,Q^2)
\,,\nn\\
& \Delta G (x,Q^2)=\delta(1-x) 
+ \sum_{n=1} \Big(\frac{\alpha_s(Q^2)}{2\pi} \Big)^n\Delta G^{(n)} (x,Q^2)
\,,\nn\\
& L_q (x,Q^2)=2n_f\delta(1-x) 
+ \sum_{n=1} \Big(\frac{\alpha_s(Q^2)}{2\pi}\Big)^nL_q^{(n)} (x,Q^2)
\,,\nn\\
& L_g (x,Q^2)=\delta(1-x) 
+ \sum_{n=1} \Big(\frac{\alpha_s(Q^2)}{2\pi} \Big)^n L_g^{(n)} (x,Q^2)
\,,\nn\\
& \Delta P_{ij} (x,Q^2)=\sum_{n=1} \Big(\frac{\alpha_s(Q^2)}{2\pi}\Big)^n 
\Delta P_{ij}^{(n)} (x,Q^2)
\,,\qquad i,j=q,g
\,,
\end{align}
and similarly for the other kernels.
Notice the prefactor $2n_f$ for the quark helicity and OAM distributions at LO, coming from the sum over all quark and antiquark flavors in their definitions in \eqref{eq:definitionsQCD}.
With these results we can easily check that the spin sum rule is indeed scale invariant:
\begin{align}
&\frac{d}{d\ln Q^2}\left( 
\frac{1}{2}\Delta\Sigma(Q^2) 
+ \Delta G(Q^2) 
+ L_q(Q^2) 
+ L_g(Q^2) \right) =
\nn\\
&\frac{\alpha_s(Q^2)}{2\pi}\sum_{i,j=q,g} 
\left( 
\Delta P^{(1)}_{ij}(\frac{1}{2}\delta_{iq}+\delta_{ig})
+ \Omega\hat P^{(1)}_{ij} 
+ \Omega P^{(1)}_{ij} 
\right)(2n_f\delta_{jq}+\delta_{jg}) 
+ O(\alpha_s^2)
= 0
\,,
\end{align}
where the coefficients $\Delta P_{ij}^{(1)}$, $\Omega\hat P_{ij}^{(1)}$ and $\Omega P_{ij}^{(1)}$ stand for the integrated splitting kernels which can be found in the appendix.

\section{Extension of the spin sum rule with QED effects}

The main goal of this short note is the inclusion of QED effects in the spin sum rule.
For this, the spin sum rule needs to be extended as
\begin{align}
\frac{1}{2}\Delta\Sigma(Q^2) 
+ \Delta G(Q^2) 
+ \frac{1}{2}\Delta l(Q^2) 
+ \Delta \gamma(Q^2) 
+ L_q(Q^2) 
+ L_g(Q^2) 
+ L_l(Q^2) 
+ L_\gamma(Q^2)
= \frac{1}{2}
\,,
\end{align}
where the newly introduced lepton/photon helicity and OAM distributions are defined in an analogous manner as their quark/gluon counterparts:
\begin{align}
&\Delta l (Q^2) = 
\sum_{f}\int_0^1dx \ (\Delta l_f(x,Q^2)+\Delta \bar{l}_f(x,Q^2))
\,,\nn\\
&\Delta \gamma(Q^2)=
\int_0^1dx \ \Delta \gamma(x,Q^2)
\,,\nn\\
& L_{l}(Q^2)=
 \sum_f \int_0^1dx \ (L_{f}(x,Q^2)+\bar{L}_f(x,Q^2))
\,,\nn\\
& L_\gamma(Q^2)=
\int_0^1 dx \ L_\gamma(x,Q^2)
\,.
\end{align}
The operators defining these distributions are completely analogous to their counterparts in QCD.
In addition, quark/gluon distributions need to be dressed with photon gauge links (see e.g. \cite{Bacchetta:2018dcq} for the case of transverse-momentum-dependent distributions), and lepton/photon distributions with gluon gauge links, in order to guarantee the full gauge-invariance of all distributions in QCD$\times$QED.

The extension of the nucleon spin sum rule is not surprising if one considers its origin (see e.g. \cite{Ji:2020ena}).
Instead of obtaining the spin sum rule starting from the angular-momentum operator in QCD, one just needs to start from the analogous operator in QCD$\times$QED, which then implies the inclusion of lepton and photon distributions.
A similar extension arises in the momentum sum rule of the nucleon, for instance, with the inclusion of a photon parton distribution function when QED evolution effects are considered (see e.g. \cite{Martin:2004dh,Roth:2004ti}).

The evolution equations get now extended as well:
\begin{align}
\label{eq:evolutionQED}
\frac{d}{d\ln Q^2} \left(\begin{matrix} 
\Delta \Sigma_U (Q^2) \\ 
\Delta \Sigma_D (Q^2) \\ 
\Delta G (Q^2) \\
\Delta l (Q^2) \\
\Delta \gamma (Q^2) \\ 
L_U(Q^2) \\
L_D(Q^2) \\
L_g(Q^2) \\
L_l(Q^2) \\
L_{\gamma}(Q^2)
\end{matrix}\right) = 
\int^1_0 dx \int_x^1 \frac{dz}{z} 
\left(\begin{matrix} 
\Delta P(z,Q^2) &  \Delta \hat P(z,Q^2) \\  
\Omega \hat P(z,Q^2) & \Omega P(z,Q^2) 
\end{matrix}\right) 
\left(\begin{matrix}   
\Delta \Sigma_U (\tfrac{x}{z},Q^2) \\
\Delta \Sigma_D (\tfrac{x}{z},Q^2) \\ 
\Delta G (\tfrac{x}{z},Q^2) \\
\Delta l (\tfrac{x}{z},Q^2) \\
\Delta \gamma (\tfrac{x}{z},Q^2) \\
L_U(\tfrac{x}{z},Q^2) \\
L_D(\tfrac{x}{z},Q^2) \\
L_g(\tfrac{x}{z},Q^2) \\
L_l(\tfrac{x}{z},Q^2) \\
L_\gamma(\tfrac{x}{z},Q^2) \\
\end{matrix}\right)
\,,
\end{align}
where the elements of the evolution matrix become a $5\times5$ matrices with again $\Delta\hat P=0$ to all orders.
We are forced to split the quark distributions into ``U-type'' and ``D-type'', to account for the electric charge dependent interactions in QED,
\begin{align}
\label{eq:expansionsQED}
\Delta \Sigma(x,Q^2) &=
\Delta \Sigma_{U}(x,Q^2) 
+ \Delta \Sigma_{D}(x,Q^2)
\,,\nn\\
\Delta \Sigma_{U(D)} (x,Q^2) &=
n_f \delta(1-x) + \sum_{n,m=0} 
\Big(\frac{\alpha_s(Q^2)}{2\pi}\Big)^n 
\Big(\frac{\alpha(Q^2)}{2\pi}\Big)^m 
\Delta\Sigma_{U(D)}^{(n,m)} (x,Q^2)
\,,
\end{align}
and similarly for $L_U$ and $L_D$.
Notice the factor $n_f$ at LO, as compared to the $2n_f$ factor in \eqref{eq:expansionsQCD}, which comes from the fact that U-type (D-type) distributions are defined as a sum only over U-type (D-type) quarks.
Notice also that in this case we need to perform the double expansion in the couplings $\as(Q^2)$ and $\a(Q^2)$.

We are now ready to check the scale-independence of the spin sum rule at LO in both QCD and QED.
First, we easily see that the QCD kernels associated with the lepton and photon distributions are all zero at LO:
\begin{align}
\Delta P_{ij}^{(1,0)}(x,Q^2) = 0
\,,\quad
i=l,\g\,,\quad j=U,D,g,l,\g
\,,\nn\\
\Delta P_{ij}^{(1,0)}(x,Q^2) = 0
\,,\quad
i=U,D,g,l,\g\,,\quad j=l,\g
\,,
\end{align}
and similarly for the analogous splitting kernels of $\Omega\hat P$ and $\Omega P$.
We also have the following simple relations between the kernels of \eqref{eq:evolutionQED} and the ones in the appendix:
\begin{align}
\Delta P_{ii}^{(1,0)}(x,Q^2) &=
\Delta P_{qq}^{(1)}(x,Q^2)
\,,\quad i=U,D\,,\nn\\
\Delta P_{ig}^{(1,0)}(x,Q^2) &=
\frac{1}{2}\Delta P_{qg}^{(1)}(x,Q^2)
\,,\quad i=U,D\,,\nn\\
\Delta P_{gj}^{(1,0)}(x,Q^2) &=
\Delta P_{gq}^{(1)}(x,Q^2)
\,,\quad j=U,D
\,,\nn\\
\Delta P_{UD}^{(1,0)}(x,Q^2) &= \Delta P_{DU}^{(1,0)}(x,Q^2) 
= 0
\,,\nn\\
\Delta P_{gg}^{(1,0)}(x,Q^2) &= \Delta P_{gg}^{(1)}(x,Q^2)
\,,
\end{align}
and similarly for the analogous splitting kernels of $\Omega\hat P$ and $\Omega P$.
Thus, we know all kernels at LO in QCD.

On the other hand, we need to calculate the kernels at LO in QED for all the distributions.
For this we use the following LO recipe (see also e.g. \cite{Bacchetta:2018dcq,deFlorian:2015ujt,deFlorian:2016gvk}) to obtain them from the known analogous kernels in QCD (see the appendix):
\begin{align}
C_F &\longrightarrow Q_i^2
\,,\nn\\
C_A &\longrightarrow  0
\,,\nn\\
T_R &\longrightarrow 1
\,,\nn\\
2n_f &\longrightarrow \sum_{i=q,\bar q,l,\bar l} N_{C,i} Q_i^2 =
2N_C \frac{n_f}{2} (Q_U^2+Q_D^2) + 2n_l
\,,
\end{align}
where the last replacement is to be used to translate a quark loop in QCD to a quark and lepton loop in QED.
The factors $2n_f$ appearing in the kernels due to the definition of the distributions as a sum over flavors, are obviously not to be treated this way.
With this recipe it is easy to see that quite some of them are zero, being left only with the following integrated non-zero kernels for $\Delta P$,
\begin{align}
&\left.\begin{aligned}
&\Delta P^{(0,1)}_{ii}= Q^2_i \int^1_0 dx \left(\frac{1+x^2}{(1-x)_+} + \frac{3}{2}\delta(1-x) \right) = 0 
\\
&\Delta P^{(0,1)}_{i\gamma}= n_f N_{C}  Q^2_i \int^1_0 \ dx \ (2x-1) = 0
\\
&\Delta P^{(0,1)}_{\gamma i}= Q^2_i \int^1_0 \ dx \ (2-x) = \frac{3}{2}Q^2_i 
\end{aligned} \ri\},\quad i=U,D
\,,\nn\\
&\Delta P^{(0,1)}_{\gamma \gamma}=  \int^1_0 dx \  \frac{\hat\beta^{(0,1)}}{2}\delta(x-1) = \frac{\hat\beta^{(0,1)}}{2}
= \frac{1}{2}\Big(
- \frac{4}{3} \Big(N_C\frac{n_f}{2} (Q_U^2+Q_D^2) + n_l
\Big)\Big) 
\,,\nn\\
&\Delta P^{(0,1)}_{ll}= Q^2_l \int^1_0 dx \left(\frac{1+x^2}{(1-x)_+} + \frac{3}{2}\delta(1-x) \right) = 0
\,,\nn\\
&\Delta P^{(0,1)}_{l\gamma}= 2n_l  Q^2_l \int^1_0 \ dx \ (2x-1) = 0
\,,\nn\\
&\Delta P^{(0,1)}_{\gamma l}= Q^2_l \int^1_0 \ dx \ (2-x) = \frac{3}{2}Q^2_l
\,,
\end{align}
for $\Omega\hat P$,
\begin{align}
&\left.\begin{aligned}
&\Omega \hat{P}^{(0,1)}_{ii} = Q^2_i \int^1_0 dx \  (x^2-1) = -\frac{2}{3}Q^2_i
\\
&\Omega \hat{P}^{(0,1)}_{i\gamma} =n_f  N_{C}  Q^2_i \int^1_0 dx \ (1-x)(1-2x+2x^2) = \frac{1}{3}n_f  N_{C}  Q^2_i
\\
&\Omega \hat{P}^{(0,1)}_{\gamma i} = Q^2_i \int^1_0 dx \ (x-1)(-x+2) = - \frac{5}{6}Q^2_i
\end{aligned}\ri\},\quad i=U,D
\,,\nn\\
&\Omega \hat{P}^{(0,1)}_{\gamma \gamma} = 0
\,,\nn\\
&\Omega \hat{P}^{(0,1)}_{ll} = Q^2_l \int^1_0 dx \  (x^2-1) = -\frac{2}{3}Q^2_l
\,,\nn\\
&\Omega \hat{P}^{(0,1)}_{l\gamma} =2n_l  Q^2_l \int^1_0 dx \ (1-x)(1-2x+2x^2) = \frac{2}{3}n_l
\,,\nn\\
&\Omega \hat{P}^{(0,1)}_{\gamma l} = Q^2_l \int^1_0 dx \ (x-1)(-x+2) = - \frac{5}{6}Q^2_l
\,,
\end{align}
and for $\Omega P$,
\begin{align}
&
\left.\begin{aligned}
&\Omega{P}^{(0,1)}_{ii} = Q^2_i \int^1_0 dx \left(\frac{x(1+x^2)}{(1-x)_+} + \frac{3}{2}\delta(1-x) \right) = -\frac{4}{3}Q^2_i
\\
&\Omega{P}^{(0,1)}_{i\gamma} = n_f  N_{C}  Q^2_i \int^1_0 dx \ x(x^2+(1-x)^2) = \frac{1}{3} n_f  N_{C}  Q^2_i 
\\
&\Omega{P}^{(0,1)}_{\gamma i} = Q^2_i \int^1_0 dx \ (1+(1-x)^2) = \frac{4}{3}Q^2_i 
\end{aligned}\right\},\quad i=U,D
\,,\nn\\
&\Omega{P}^{(0,1)}_{\gamma \gamma} =  \int^1_0 dx \ \frac{\hat\beta^{(0,1)}}{2}\delta(x-1) = \frac{\hat\beta^{(0,1)}}{2}
= \frac{1}{2}\Big(
- \frac{4}{3} \Big(N_C\frac{n_f}{2} (Q_U^2+Q_D^2) + n_l
\Big)\Big)
\,,\nn\\
&\Omega{P}^{(0,1)}_{ll} = Q^2_l \int^1_0 dx \left(\frac{x(1+x^2)}{(1-x)_+} + \frac{3}{2}\delta(1-x) \right) = -\frac{4}{3}Q^2_l 
\,,\nn\\
&\Omega{P}^{(0,1)}_{\gamma l} = Q^2_l \int^1_0 dx \ (1+(1-x)^2) = \frac{4}{3}Q^2_l
\,,\nn\\
&\Omega{P}^{(0,1)}_{l\gamma} =2 n_l  Q^2_l \int^1_0 dx \ x(x^2+(1-x)^2) = \frac{2}{3} n_l
\,.
\end{align}

With all these results, we can finally check the scale-independence of the spin sum rule at LO in both QCD and QED:
\begin{align}
&\frac{d}{d\ln Q^2}\left( 
\frac{1}{2}\Delta\Sigma(Q^2) 
+ \Delta G(Q^2) 
+ \frac{1}{2}\Delta l(Q^2) 
+ \Delta \gamma(Q^2) 
+ L_q(Q^2) 
+ L_g(Q^2) 
+ L_l(Q^2) 
+ L_\gamma(Q^2) \right) = 
\nn\\
&
\Bigg[\frac{\alpha_s(Q^2)}{2\pi}
\sum_{i,j=U,D,g,\gamma,l} 
\left( \Delta P^{(1,0)}_{ij} 
\Big(\frac{1}{2}\delta_{iU}+\frac{1}{2}\delta_{iD}+\delta_{ig}
+\frac{1}{2}\delta_{il}+\delta_{i\gamma}\Big)
+ \Omega\hat{P}^{(1,0)}_{ij} + \Omega{P}^{(1,0)}_{ij} \right)
\nonumber\\
&\,\,\,
+ \frac{\alpha(Q^2)}{2\pi} 
\sum_{i,j = U,D,g,\gamma,l} 
\left( \Delta P^{(0,1)}_{ij} 
\Big(\frac{1}{2}\delta_{iU}+\frac{1}{2}\delta_{iD}+\delta_{ig}
+\frac{1}{2}\delta_{il}+\delta_{i\gamma}\Big)
+ \Omega\hat{P}^{(0,1)}_{ij} + \Omega{P}^{(0,1)}_{ij} \right)\Bigg]
\nn\\
&\times
\Big(n_f\delta_{jU}
+n_f\delta_{jD}
+\delta_{jg}
+2n_l\delta_{jl}
+\delta_{j\gamma}\Big)
+ O(\alpha^2,\alpha_s^2, \alpha \alpha_s)= 0
\,.
\end{align}

\section{Conclusions}
In this short note the inclusion of QED effects on the nucleon spin sum rule have been considered.
To that end, new lepton/photon helicity and orbital angular momentum (OAM) distributions have been introduced, as well as all the needed new splitting functions at leading order (LO) obtained: on one hand the new splitting functions at LO in QED for the quark and gluon distributions, and on the other the new splitting functions at LO in both QED and QCD of the newly introduced lepton/photon distributions.
Putting all together, the expected scale-invariance of the extended nucleon spin sum rule at LO has been explicitly checked.

This formalism will allow in the future the phenomenological determination of the contribution of lepton/photon helicity and OAM distributions to the nucleon spin.

\textbf{Acknowledgements.}
M.G.E. is supported by the Spanish Ministry grant PID2019-106080GB-C21.

\appendix

\section{Evolution kernels in QCD}

Evolution kernels of the quark/gluon helicity and OAM distributions at LO in QCD, integrated over $x$:
\begin{align}
&\Delta P^{(1)}_{qq}=C_F \int^1_0 dx \left(\frac{1+x^2}{(1-x)_+} + \frac{3}{2}\delta(1-x) \right) = 0
\,,\nn\\
&\Delta P^{(1)}_{qg}= 2 \ n_f \ T_R \ \int^1_0 \ dx \ (2x-1) = 0
\,,\nn\\
&\Delta P^{(1)}_{gq}= C_F \int^1_0 \ dx \ (2-x) = \frac{3}{2}C_F
\,,\nn\\
&\Delta P^{(1)}_{gg}= 2 \ C_A \int^1_0 dx \left(\frac{1}{(1-x)_+}-2x+1\right) +  \frac{\beta^{(1)}}{2}\delta(x-1) = \frac{\beta^{(1)}}{2}
= \frac{1}{2}\Big(
\frac{11}{3} C_A - \frac{4}{3} T_R n_f\Big) 
\,,
\end{align}

\begin{align}
\Omega \hat{P}^{(1)}_{qq} &= C_F \int^1_0 dx \  (x^2-1) = -\frac{2}{3}C_F
\,,\nn\\
\Omega \hat{P}^{(1)}_{qg} &=2 \ n_f \ T_R \int^1_0 dx \ (1-x)(1-2x+2x^2) = \frac{2}{3} n_f \ T_R
\,,\nn\\
\Omega \hat{P}^{(1)}_{gq} &= C_F \int^1_0 dx \ (x-1)(-x+2) = - \frac{5}{6}C_F
\,,\nn\\
\Omega \hat{P}^{(1)}_{gg} &= 2 \ C_A \int^1_0 dx \ (x-1)(x^2-x+2) = -\frac{11}{6}C_A
\,,
\end{align}

\begin{align}
\Omega P^{(1)}_{qq} &= C_F \int^1_0 dx \left(\frac{x(1+x^2)}{(1-x)_+} + \frac{3}{2}\delta(1-x) \right) = -\frac{4}{3}C_F 
\,,\nn\\
\Omega P^{(1)}_{qg} &= 2 \ n_f \ T_R \int^1_0 dx \ x(x^2+(1-x)^2) = \frac{2}{3} n_f \ T_R 
\,,\nn\\
\Omega P^{(1)}_{gq} &= C_F \int^1_0 dx \ (1+(1-x)^2) = \frac{4}{3}C_F 
\,,\nn\\
\Omega P^{(1)}_{gg} &= 2 \ C_A \int^1_0 dx \ \frac{(x^2-x+1)^2}{(1-x)_+} +  \frac{\beta^{(1)}}{2}\delta(x-1) = \frac{\beta^{(1)}}{2} - \frac{11}{6}C_A
\,.
\end{align}

\bibliographystyle{utphys}
\bibliography{references}

\providecommand{\href}[2]{#2}\begingroup\raggedright\begin{thebibliography}{10}

\bibitem{Jaffe:1989jz}
R.~L. Jaffe and A.~Manohar, ``{The G(1) Problem: Fact and Fantasy on the Spin
  of the Proton},'' \href{http://dx.doi.org/10.1016/0550-3213(90)90506-9}{{\em
  Nucl. Phys. B} {\bfseries 337} (1990) 509--546}.

\bibitem{Ahmed:1976ee}
M.~A. Ahmed and G.~G. Ross, ``{Polarized Lepton - Hadron Scattering in
  Asymptotically Free Gauge Theories},''
  \href{http://dx.doi.org/10.1016/0550-3213(76)90328-X}{{\em Nucl. Phys. B}
  {\bfseries 111} (1976) 441--460}.

\bibitem{Altarelli:1977zs}
G.~Altarelli and G.~Parisi, ``{Asymptotic Freedom in Parton Language},''
  \href{http://dx.doi.org/10.1016/0550-3213(77)90384-4}{{\em Nucl. Phys. B}
  {\bfseries 126} (1977) 298--318}.

\bibitem{Kodaira:1979pa}
J.~Kodaira, ``{QCD Higher Order Effects in Polarized Electroproduction: Flavor
  Singlet Coefficient Functions},''
  \href{http://dx.doi.org/10.1016/0550-3213(80)90310-7}{{\em Nucl. Phys. B}
  {\bfseries 165} (1980) 129--140}.

\bibitem{Mertig:1995ny}
R.~Mertig and W.~L. van Neerven, ``{The Calculation of the two loop spin
  splitting functions P(ij)(1)(x)},''
  \href{http://dx.doi.org/10.1007/s002880050138}{{\em Z. Phys. C} {\bfseries
  70} (1996) 637--654}, \href{http://arxiv.org/abs/hep-ph/9506451}{{\ttfamily
  arXiv:hep-ph/9506451}}.

\bibitem{Vogelsang:1995vh}
W.~Vogelsang, ``{A Rederivation of the spin dependent next-to-leading order
  splitting functions},''
  \href{http://dx.doi.org/10.1103/PhysRevD.54.2023}{{\em Phys. Rev. D}
  {\bfseries 54} (1996) 2023--2029},
  \href{http://arxiv.org/abs/hep-ph/9512218}{{\ttfamily arXiv:hep-ph/9512218}}.

\bibitem{Vogelsang:1996im}
W.~Vogelsang, ``{The Spin dependent two loop splitting functions},''
  \href{http://dx.doi.org/10.1016/0550-3213(96)00306-9}{{\em Nucl. Phys. B}
  {\bfseries 475} (1996) 47--72},
  \href{http://arxiv.org/abs/hep-ph/9603366}{{\ttfamily arXiv:hep-ph/9603366}}.

\bibitem{Vogt:2008yw}
A.~Vogt, S.~Moch, M.~Rogal, and J.~A.~M. Vermaseren, ``{Towards the NNLO
  evolution of polarised parton distributions},''
  \href{http://dx.doi.org/10.1016/j.nuclphysbps.2008.09.097}{{\em Nucl. Phys. B
  Proc. Suppl.} {\bfseries 183} (2008) 155--161},
  \href{http://arxiv.org/abs/0807.1238}{{\ttfamily arXiv:0807.1238 [hep-ph]}}.

\bibitem{Moch:2014sna}
S.~Moch, J.~A.~M. Vermaseren, and A.~Vogt, ``{The Three-Loop Splitting
  Functions in QCD: The Helicity-Dependent Case},''
  \href{http://dx.doi.org/10.1016/j.nuclphysb.2014.10.016}{{\em Nucl. Phys. B}
  {\bfseries 889} (2014) 351--400},
  \href{http://arxiv.org/abs/1409.5131}{{\ttfamily arXiv:1409.5131 [hep-ph]}}.

\bibitem{Moch:2015usa}
S.~Moch, J.~A.~M. Vermaseren, and A.~Vogt, ``{On \ensuremath{\gamma}5 in
  higher-order QCD calculations and the NNLO evolution of the polarized valence
  distribution},'' \href{http://dx.doi.org/10.1016/j.physletb.2015.07.027}{{\em
  Phys. Lett. B} {\bfseries 748} (2015) 432--438},
  \href{http://arxiv.org/abs/1506.04517}{{\ttfamily arXiv:1506.04517
  [hep-ph]}}.

\bibitem{Ji:1995cu}
X.-D. Ji, J.~Tang, and P.~Hoodbhoy, ``{The spin structure of the nucleon in the
  asymptotic limit},'' \href{http://dx.doi.org/10.1103/PhysRevLett.76.740}{{\em
  Phys. Rev. Lett.} {\bfseries 76} (1996) 740--743},
  \href{http://arxiv.org/abs/hep-ph/9510304}{{\ttfamily arXiv:hep-ph/9510304}}.

\bibitem{Hagler:1998kg}
P.~Hagler and A.~Schafer, ``{Evolution equations for higher moments of angular
  momentum distributions},''
  \href{http://dx.doi.org/10.1016/S0370-2693(98)00414-6}{{\em Phys. Lett. B}
  {\bfseries 430} (1998) 179--185},
  \href{http://arxiv.org/abs/hep-ph/9802362}{{\ttfamily arXiv:hep-ph/9802362}}.

\bibitem{Harindranath:1998ve}
A.~Harindranath and R.~Kundu, ``{On Orbital angular momentum in deep inelastic
  scattering},'' \href{http://dx.doi.org/10.1103/PhysRevD.59.116013}{{\em Phys.
  Rev. D} {\bfseries 59} (1999) 116013},
  \href{http://arxiv.org/abs/hep-ph/9802406}{{\ttfamily arXiv:hep-ph/9802406}}.

\bibitem{Hatta:2018itc}
Y.~Hatta and D.-J. Yang, ``{On the small-$x$ behavior of the orbital angular
  momentum distributions in QCD},''
  \href{http://dx.doi.org/10.1016/j.physletb.2018.03.081}{{\em Phys. Lett. B}
  {\bfseries 781} (2018) 213--219},
  \href{http://arxiv.org/abs/1802.02716}{{\ttfamily arXiv:1802.02716
  [hep-ph]}}.

\bibitem{Bacchetta:2018dcq}
A.~Bacchetta and M.~G. Echevarria, ``{QCD$\times$QED evolution of TMDs},''
  \href{http://dx.doi.org/10.1016/j.physletb.2018.11.019}{{\em Phys. Lett. B}
  {\bfseries 788} (2019) 280--287},
  \href{http://arxiv.org/abs/1810.02297}{{\ttfamily arXiv:1810.02297
  [hep-ph]}}.

\bibitem{Ji:2020ena}
X.~Ji, F.~Yuan, and Y.~Zhao, ``{What we know and what we don\textquoteright{}t
  know about the proton spin after 30 years},''
  \href{http://dx.doi.org/10.1038/s42254-020-00248-4}{{\em Nature Rev. Phys.}
  {\bfseries 3} no.~1, (2021) 27--38},
  \href{http://arxiv.org/abs/2009.01291}{{\ttfamily arXiv:2009.01291
  [hep-ph]}}.

\bibitem{Martin:2004dh}
A.~D. Martin, R.~G. Roberts, W.~J. Stirling, and R.~S. Thorne, ``{Parton
  distributions incorporating QED contributions},''
  \href{http://dx.doi.org/10.1140/epjc/s2004-02088-7}{{\em Eur. Phys. J. C}
  {\bfseries 39} (2005) 155--161},
  \href{http://arxiv.org/abs/hep-ph/0411040}{{\ttfamily arXiv:hep-ph/0411040}}.

\bibitem{Roth:2004ti}
M.~Roth and S.~Weinzierl, ``{QED corrections to the evolution of parton
  distributions},''
  \href{http://dx.doi.org/10.1016/j.physletb.2004.04.009}{{\em Phys. Lett. B}
  {\bfseries 590} (2004) 190--198},
  \href{http://arxiv.org/abs/hep-ph/0403200}{{\ttfamily arXiv:hep-ph/0403200}}.

\bibitem{deFlorian:2015ujt}
D.~de~Florian, G.~F.~R. Sborlini, and G.~Rodrigo, ``{QED corrections to the
  Altarelli\textendash{}Parisi splitting functions},''
  \href{http://dx.doi.org/10.1140/epjc/s10052-016-4131-8}{{\em Eur. Phys. J. C}
  {\bfseries 76} no.~5, (2016) 282},
  \href{http://arxiv.org/abs/1512.00612}{{\ttfamily arXiv:1512.00612
  [hep-ph]}}.

\bibitem{deFlorian:2016gvk}
D.~de~Florian, G.~F.~R. Sborlini, and G.~Rodrigo, ``{Two-loop QED corrections
  to the Altarelli-Parisi splitting functions},''
  \href{http://dx.doi.org/10.1007/JHEP10(2016)056}{{\em JHEP} {\bfseries 10}
  (2016) 056}, \href{http://arxiv.org/abs/1606.02887}{{\ttfamily
  arXiv:1606.02887 [hep-ph]}}.

\end{thebibliography}\endgroup

\end{document}